\documentclass[aps,prc,preprint,superscriptaddress,showpacs]{revtex4}

\usepackage[dvips]{graphicx}
\usepackage{amsmath,amssymb,times}

\newcommand{\bequ}{\begin{equation}}
\newcommand{\eequ}{\end{equation}}
\newcommand{\bea}{\begin{eqnarray}}
\newcommand{\eea}{\end{eqnarray}}

\DeclareSymbolFont{boldletters}{OML}{cmm} {b}{it}
\DeclareSymbolFontAlphabet{\mathbit}{boldletters}
\DeclareMathSymbol{\alpha}{\mathalpha}{letters}{"0B}
\DeclareMathSymbol{\beta}{\mathalpha}{letters}{"0C}
\DeclareMathSymbol{\gamma}{\mathalpha}{letters}{"0D}
\DeclareMathSymbol{\delta}{\mathalpha}{letters}{"0E}
\DeclareMathSymbol{\epsilon}{\mathalpha}{letters}{"0F}
\DeclareMathSymbol{\zeta}{\mathalpha}{letters}{"10}
\DeclareMathSymbol{\eta}{\mathalpha}{letters}{"11}
\DeclareMathSymbol{\theta}{\mathalpha}{letters}{"12}
\DeclareMathSymbol{\iota}{\mathalpha}{letters}{"13}
\DeclareMathSymbol{\kappa}{\mathalpha}{letters}{"14}
\DeclareMathSymbol{\lambda}{\mathalpha}{letters}{"15}
\DeclareMathSymbol{\mu}{\mathalpha}{letters}{"16}
\DeclareMathSymbol{\nu}{\mathalpha}{letters}{"17}
\DeclareMathSymbol{\xi}{\mathalpha}{letters}{"18}
\DeclareMathSymbol{\pi}{\mathalpha}{letters}{"19}
\DeclareMathSymbol{\rho}{\mathalpha}{letters}{"1A}
\DeclareMathSymbol{\sigma}{\mathalpha}{letters}{"1B}
\DeclareMathSymbol{\tau}{\mathalpha}{letters}{"1C}
\DeclareMathSymbol{\upsilon}{\mathalpha}{letters}{"1D}
\DeclareMathSymbol{\phi}{\mathalpha}{letters}{"1E}
\DeclareMathSymbol{\chi}{\mathalpha}{letters}{"1F}
\DeclareMathSymbol{\psi}{\mathalpha}{letters}{"20}
\DeclareMathSymbol{\omega}{\mathalpha}{letters}{"21}
\DeclareMathSymbol{\varepsilon}{\mathalpha}{letters}{"22}
\DeclareMathSymbol{\vartheta}{\mathalpha}{letters}{"23}
\DeclareMathSymbol{\varpi}{\mathalpha}{letters}{"24}
\DeclareMathSymbol{\varrho}{\mathalpha}{letters}{"25}
\DeclareMathSymbol{\varsigma}{\mathalpha}{letters}{"26}
\DeclareMathSymbol{\varphi}{\mathalpha}{letters}{"27}
\DeclareMathSymbol{\Gamma}{\mathalpha}{letters}{"00}
\DeclareMathSymbol{\Delta}{\mathalpha}{letters}{"01}
\DeclareMathSymbol{\Theta}{\mathalpha}{letters}{"02}
\DeclareMathSymbol{\Lambda}{\mathalpha}{letters}{"03}
\DeclareMathSymbol{\Xi}{\mathalpha}{letters}{"04}
\DeclareMathSymbol{\Pi}{\mathalpha}{letters}{"05}
\DeclareMathSymbol{\Sigma}{\mathalpha}{letters}{"06}
\DeclareMathSymbol{\Upsilon}{\mathalpha}{letters}{"07}
\DeclareMathSymbol{\Phi}{\mathalpha}{letters}{"08}
\DeclareMathSymbol{\Psi}{\mathalpha}{letters}{"09}
\DeclareMathSymbol{\Omega}{\mathalpha}{letters}{"0A}

\begin{document}
\preprint{SAGA-HE-267}
\title{Theta vacuum effects on QCD phase diagram}

\author{Yuji Sakai}
\email[]{sakai@phys.kyushu-u.ac.jp}
\affiliation{Department of Physics, Graduate School of Sciences, Kyushu University,
             Fukuoka 812-8581, Japan}

\author{Hiroaki Kouno}
\email[]{kounoh@cc.saga-u.ac.jp}
\affiliation{Department of Physics, Saga University,
             Saga 840-8502, Japan}

\author{Takahiro Sasaki}
\email[]{sasaki@phys.kyushu-u.ac.jp}
\affiliation{Department of Physics, Graduate School of Sciences, Kyushu University,
             Fukuoka 812-8581, Japan}

\author{Masanobu Yahiro}
\email[]{yahiro@phys.kyushu-u.ac.jp}
\affiliation{Department of Physics, Graduate School of Sciences, Kyushu University,
             Fukuoka 812-8581, Japan}

\date{\today}

\begin{abstract}
Theta vacuum effects on  the QCD phase structure 
in the $\mu$-$T$ plane are studied  
by using the Polyakov-loop extended Nambu-Jona-Lasinio model and 
its extension, 
where $\mu$ is the quark chemical potential and $T$ is temperature, 
respectively.  
As the parameter $\theta$ of the theta vacuum increases, 
the chiral transition becomes stronger. 
For large $\theta$, it eventually becomes first order even at zero $\mu$. 
\end{abstract}

\pacs{11.10.Wx, 12.38.Mh, 11.30.Rd, 12.40.-y}
\maketitle

\section{Introduction}
\label{Introduction}

Violations of parity ($P$), charge conjugation ($C$) and 
charge-parity symmetries ($CP$) are important subjects in particle and nuclear physics. 
For example, the strong $CP$ problem is a long-standing puzzle; 
see for example Ref.~\cite{Vicari} for a review of this problem. 
Lorentz and gauge invariance allow the Quantum Chromodynamics (QCD) 
action to have a term 
\bea
   {\cal L}_{\theta}=\theta \frac{g^2}{64\pi^2}\epsilon^{\mu\nu\sigma\rho}
   F^{a}_{\mu\nu}F^{a}_{\sigma\rho} 
\eea
of the topological charge, where 
$F^{a}_{\mu\nu}$ is the field strength of gluon. The parameter  
$\theta$ can take any arbitrary value between $-\pi$ and $\pi$, where 
$\theta=-\pi$ is identical with $\theta=\pi$. 
Nevertheless, experiment indicates 
$|\theta| < 3 \times 10^{-10}$~\cite{Baker,Ohta}. 
Since $\theta$ is $P$-odd ($CP$-odd), 
$P$ ($CP$) is then preserved for $\theta=0$ and $\pm \pi$, but 
explicitly broken for other $\theta$. 
Why is $\theta$ so small ? This is the so-called strong $CP$ problem. 

For zero temperature ($T$) and zero quark-chemical potential ($\mu$), 
$P$ is conserved at $\theta=0$, as Vafa and Witten showed~\cite{VW}. 
Meanwhile, $P$ is spontaneously broken 
at $\theta =\pi$, as Dashen~\cite{Dashen} 
and Witten~\cite{Witten} pointed out. 
This is the so-called Dashen phenomena. 
Since the spontaneous $P$ violation is a nonperturbative phenomenon, 
the phenomenon was so far studied mainly with the effective model 
such as the chiral perturbation theory~\cite{VV,Smilga,Tytgat,ALS,Creutz,MZ}, the Nambu-Jona-Lasinio (NJL) model~\cite{FIK,Boer,Boomsma}  
and the Polyakov-loop extended Nambu-Jona-Lasinio (PNJL) 
model~\cite{Kouno_CP}. 

For $T$ higher than the QCD scale $\Lambda_{\mathrm QCD}$, 
there is a possibility that a finite $\theta$, depending 
on spacetime coordinates $(t,x)$, is effectively 
induced, 
since sphalerons are so activated as to jump over the potential 
barrier between the different degenerate ground states~\cite{MMS}. 
If so, $P$ and $CP$ symmetries can be violated locally 
in high-energy heavy-ion collisions or the early universe 
at $T \approx \Lambda_{\mathrm QCD}$. 
Actually, it is argued in Refs.~\cite{MZ,KZ} that $\theta$ may be of order one 
during the QCD phase transition in the early universe, 
while it vanishes at the present epoch
~\cite{Peccei,Dine,Zhitnitsky,Shifman,Kim}. 
This finite $\theta$ may be a new source of very large $CP$ violation 
in the Universe and may be a crucial missing element 
for solving the puzzle of baryogenesis. 

Furthermore, this effective $\theta (t,x)$ deviates 
the total number of particles plus antiparticles with right-handed 
helicity from that with left-handed helicity. 
The magnetic field, formed in the early stage of heavy-ion collision, 
will lift the degeneracy in spin depending on the charge of particle. 
As a consequence of this fact, 
an electromagnetic current is generated along the magnetic field, 
since particles with right-handed helicity move opposite 
to antiparticles with right-handed helicity. 
This is the so-called chiral magnetic effect 
(CME)~\cite{Kharzeev,KZ,FKW,Fukushima3}. 
CME may explain the charge separations 
observed in the recent STAR results~\cite{Abelev}. 
Thus, the thermal system with nonzero $\theta$ is quite interesting.

In this letter, we study effects of the theta-vacuum on QCD phase diagram 
by using the two-flavor PNJL model~\cite{Meisinger,Fukushima,Ratti,Rossner,Schaefer,Kashiwa1,Sakai,Sakai1,Sakai2,Kouno,Sakai3,Kashiwa5,Matsumoto,Sasaki-T,Sakai5,Gatto:2010pt,Kouno_CP,Sakai_EOS} and its extension; see Ref.~\cite{Kouno_CP} 
and references therein for further information on the PNJL model. 
Particularly, our attention is focused on the two-flavor phase diagram  
in the $\mu$-$T$ plane, where $\mu$ is the quark chemical potential. 
In our previous work Ref.~\cite{Kouno_CP}, 
as a theoretical interest, we investigated spontaneous $P$ and $C$ violations 
at finite $\theta$ and imaginary $\mu$. 
As physical phenomena that may occur in the early universe or 
the high-energy heavy-ion collisions, we here examine 
the chiral and $P$ symmetry restorations at finite $T$ and $\theta$ and 
real $\mu$.

This paper is organized as follows. 
In section II, the PNJL model and its extension are explained briefly. 
In section III, the numerical results are shown. 
Section IV is devoted to summary.

\section{PNJL model}
\label{PNJL}

Pioneering work on the parity violation and its restoration in the framework 
of the NJL model~\cite{NJ1,AY,Kashiwa,FIK,Boer,Boomsma} 
was done by Fujihara, Inagaki and Kimura~\cite{FIK}. 
Boer and Boomsma studied this issue extensively~\cite{Boer,Boomsma}. 
In Ref.~\cite{Kouno_CP}, we have extended their formalism based on the NJL model to that on 
the PNJL model. 
The two-flavor ($N_{f}=2$) PNJL Lagrangian 
with the $\theta$-dependent anomaly term is 
\begin{eqnarray}
{\cal L} &=& {\bar q}(i \gamma_\nu D^\nu -m)q 
	- {\cal U}(\Phi [A],{\Phi} [A]^*,T)
\notag\\
         &+& G_1\sum_{a=0}^3\left[({\bar q}\tau_a q)^2 
               +({\bar q}i\gamma_5 \tau_a q)^2\right]
          + 8G_2\left[e^{i\theta}\det{\left(\bar{q}_{\rm R}q_{\rm L}\right)}
               +e^{-i\theta}\det{\left(\bar{q}_{\rm L}q_{\rm R}\right)}\right], \label{eq:E1}
\end{eqnarray}
where $q=(u,d)$ denotes the two-flavor quark field, 
$m$ does the current quark-mass matrix ${\rm diag}(m_u,m_d)$, 
$\tau_0$ is the $2\times 2$ unit matrix, 
$\tau_a (a=1,2,3)$ are the Pauli matrices 
and $D^\nu=\partial^\nu+iA^\nu-i\mu\delta^{\nu}_{0}$. 
The field $A^\nu$ is defined as $A^\nu=\delta^{\nu}_{0}gA^0_a{\lambda^a\over{2}}$ with the gauge field $A^\nu_a$, the Gell-Mann matrix $\lambda_a$ and the gauge coupling $g$. 
In the NJL sector, $G_1$ denotes the coupling constant of the scalar and pseudoscalar-type four-quark interaction, and 
$G_2$ is the coupling constant of the Kobayashi-Maskawa-'t Hooft determinant interaction~\cite{KMK,'t Hooft} the matrix indices of which run in the flavor space. 
The Polyakov potential ${\cal U}$, defined later in (\ref{eq:E13}), 
is a function of the Polyakov loop $\Phi$ and 
its Hermitian conjugate $\Phi^*$, 
\begin{align}
\Phi      = {1\over{N_{\rm c}}}{\rm Tr} L,~~~~
\Phi^{*}  = {1\over{N_{\rm c}}} {\rm Tr}L^\dag 
\end{align}
with
\begin{align}
L({\bf x}) = {\cal P} \exp\Bigl[
                {i\int^\beta_0 d \tau A_4({\bf x},\tau)}\Bigr],
\end{align}
where ${\cal P}$ is the path ordering and $A_4 = iA_0 $. 
In the chiral limit ($m_{\rm u}=m_{\rm d}=0$), the Lagrangian density has the exact $SU(N_f)_{\rm L} \times SU(N_f)_{\rm R}\times U(1)_{\rm v} \times SU(3)_{\rm c}$  symmetry. 
The $U(1)_{\rm A}$ symmetry is explicitly broken if $G_2\neq 0$. 
The temporal component of the gauge field is diagonal in the flavor space, 
because the color and the flavor spaces are completely separated out 
in the present case. 
In the Polyakov gauge, $L$ can be written in a diagonal form in the color space~\cite{Fukushima}: 
\begin{align}
L 
=  e^{i \beta (\phi_3 \lambda_3 + \phi_8 \lambda_8)}
= {\rm diag} (e^{i \beta \phi_a},e^{i \beta \phi_b},
e^{i \beta \phi_c} ),
\label{eq:E6}
\end{align}
where $\phi_a=\phi_3+\phi_8/\sqrt{3}$, $\phi_b=-\phi_3+\phi_8/\sqrt{3}$
and $\phi_c=-(\phi_a+\phi_b)=-2\phi_8/\sqrt{3}$. 
The Polyakov loop $\Phi$ is an exact order parameter of the spontaneous 
${\mathbb Z}_3$ symmetry breaking in the pure gauge theory.
Although the ${\mathbb Z}_3$ symmetry is not an exact one 
in the system with dynamical quarks, it may be a good indicator of 
the deconfinement phase transition. 
Therefore, we use $\Phi$ to define the deconfinement phase transition.
For simplicity, we assume below that $m_{\rm u}=m_{\rm d}=m_0$. 

We transform the quark field $q$ to the new one $q^\prime$ with 
\begin{eqnarray}
q_R=e^{i{\theta\over{4}}}q^\prime_R,~~~~~q_L
=e^{-i{\theta\over{4}}}q^\prime_L,  
\label{UAtrans}
\end{eqnarray}
in order to remove the $\theta$ dependence of the determinant interaction. 
Under this ${\rm U}(1)_{\rm A}$ transformation, 
the quark-antiquark condensates are also transformed as 
\begin{eqnarray}
\sigma &\equiv &\bar{q}q
=\cos{\left({\theta\over{2}}\right)}\sigma^\prime
+\sin{\left({\theta\over{2}}\right)}\eta^\prime ,
\nonumber\\
\eta &\equiv &\bar{q}i\gamma_5q
=-\sin{\left({\theta\over{2}}\right)}\sigma^\prime
+\cos{\left({\theta\over{2}}\right)}\eta^\prime ,
\nonumber\\
a_i &\equiv &\bar{q}\tau_iq
=\cos{\left({\theta\over{2}}\right)}a_i^\prime
+\sin{\left({\theta\over{2}}\right)}\pi_i^\prime ,
\nonumber\\
\pi_i &\equiv &\bar{q}i\tau_i \gamma_5q
=-\sin{\left({\theta\over{2}}\right)}a_i^\prime
+\cos{\left({\theta\over{2}}\right)}\pi_i^\prime , 
\label{Econdensates}
\end{eqnarray}
where $\sigma^\prime$ is defined by the same form as $\sigma$ but $q$ is 
replaced by $q^\prime$; this is the case also for other condensates 
$\eta^\prime$, $a_i^\prime$ and $\pi_i^\prime$. 
The Lagrangian density is then rewritten with $q^\prime$ as 
\begin{eqnarray}
{\cal L}  
&=& {\bar q^\prime}(i \gamma_\nu D^\nu -m_{0+}-im_{0-}\gamma_5)q^\prime 
          - {\cal U}(\Phi [A],{\Phi} [A]^*,T) 
\nonumber\\
          &+& G_1\sum_{a=0}^3\left[({\bar q^\prime }\tau_aq^\prime )^2 
               +({\bar q^\prime}i\gamma_5 \tau_aq^\prime)^2\right] 
\nonumber\\
          &+& 8G_2\left[\det{\left(\bar{q^\prime }_{\rm R}q_{\rm L}^\prime 
               \right)}
          +\det{\left(\bar{q^\prime}_{\rm L}q_{\rm R}^\prime\right)}\right] ,
          \label{eq:E1rewritten}
\end{eqnarray}
where $m_{0+}=m_0\cos{\left({\theta\over{2}}\right)}$ and 
$m_{0-}=m_0\sin{\left({\theta\over{2}}\right)}$. 
Making the mean field approximation and performing 
the path integral over the quark field, 
one can obtain the thermodynamic potential $\Omega$ (per volume) 
for finite $T$ and $\mu$: 
\begin{align}
\Omega =& -2 \int \frac{d^3{\rm p}}{(2\pi)^3}
         \Bigl[ 3 \{E_+ ({\rm p})+E_-({\rm p})\} 
         + \frac{1}{\beta}
           \ln~ [1 + 3\Phi e^{-\beta E_+^{-} }
           +3\Phi^{*}e^{-2\beta E_+^{-}}
           + e^{-3\beta E_+^{-}}]
\notag\\
       &+ \frac{1}{\beta}
           \ln~ [1 + 3\Phi e^{-\beta E_-^{-} }
           +3\Phi^{*}e^{-2\beta E_-^{-}}
           + e^{-3\beta E_-^{-}}]
        + \frac{1}{\beta}
           \ln~ [1 + 3\Phi^* e^{-\beta E_+^{+} }
           +3\Phi e^{-2\beta E_+^{+}}
           + e^{-3\beta E_+^{+}}]
         \nonumber\\ 
       & + \frac{1}{\beta}
           \ln~ [1 + 3\Phi^* e^{-\beta E_-^{+} }
           +3\Phi e^{-2\beta E_-^{+}}
           + e^{-3\beta E_-^{+}}]
         +U+{\cal U},  
\label{eq:E12} 
\end{align}
where $E_+^{\pm}=E_+({\bf p}) \pm \mu$ and 
$E_-^{\pm}=E_-({\bf p}) \pm \mu$ with 
\begin{eqnarray}
E_{\pm}&=&\sqrt{{\bf p}^2+C \pm 2\sqrt{D}}, 
\label{E12a}\\
C&=&M^2+N^2+A^2+P^2, 
\label{E12c}\\
D&=&A^2M^2+P^2N^2+2APMN\cos{\varphi}+A^2P^2\sin^2{\varphi}
\nonumber\\
&=&(M{\bf A}+N{\bf P})^2+({\bf A}\times{\bf P})^2\ge 0
\\
\label{E12e}
M&=&m_{0+}-2G_+\sigma^\prime =m_{0+}-2(G_1+G_2)\sigma^\prime ,
\label{E12M}\\
N&=&m_{0-}-2G_-\eta^\prime =m_{0-}-2(G_1-G_2)\eta^\prime , 
\label{E12N}\\
{\bf A}&=&(-2G_-a_1^\prime ,-2G_-a_2^\prime ,-2G_-a_3^\prime ), 
\label{E12Ameson}\\
{\bf P}&=&(-2G_+\pi_1^\prime ,-2G_+\pi_2^\prime ,-2G_+\pi_3^\prime ), 
\label{E12Pion}\\
A&=&\sqrt{{\bf A}\cdot{\bf A}},~~~P=\sqrt{{\bf P}\cdot{\bf P}},~~~{\bf A}\cdot{\bf P}=AP\cos{\varphi},
\label{E12absoluteangle}\\
U&=&G_+({\sigma^\prime}^2+{\pi_a^\prime}^2)+G_-({a_a^\prime}^2+{\eta^\prime}^2) .
\label{E12U}
\end{eqnarray}
In the right-hand side of \eqref{eq:E12}, only the first term diverges. 
The term is then regularized by the three-dimensional momentum
cutoff $\Lambda$~\cite{Fukushima,Ratti}. 
Following Ref.~\cite{Boer,Boomsma}, we introduce $c$ as 
$G_1=(1-c)G$ and $G_2=cG$, 
where $0\leq c\leq 0.5$ and $G>0$. 
Hence, the NJL sector has four parameters of $m_0$, $\Lambda$, $G$ and $c$. 
We put $m_0=5.5$MeV. 
The parameters $\Lambda$ and $G$ are so chosen as to reproduce 
the pion decay constant $f_\pi =93$MeV and the pion mass $m_\pi =139$MeV 
at vacuum. 
The remaining parameter $c$ is a free parameter. 
Although the exact value of $c$ is unknown, 
it is known from the analysis of the $\eta$-$\eta^\prime$ splitting 
in the three-flavor model that $c\sim 0.2$ is favorable~\cite{FBO}. 
The value $c=0.2$ has been also used in Refs.~\cite{Boer,Boomsma}. 
Therefore, we take $c=0.2$ also here. 

The PNJL model is reduced to the NJL model 
in the limit of $T=0$, 
since in \eqref{eq:E12} the Polyakov-loop potential ${\cal U}$ 
tends to zero and 
the logarithmic terms approach $N_c\Theta (\mu -E_{\pm}^{\pm})$, 
where $\Theta (x)=1$ for $x>0$ and $\Theta (x)=0$ for $x<0$, in 
the limit. 
Since $P$ breaking at $T=0$ and $\theta=\pi$ has already been 
studied in detail by the NJL model~\cite{FIK,Boer,Boomsma}, 
we here concentrate ourselves on $P$ restoration at finite $T$ and 
$\theta=\pi$. 

The Polyakov potential ${\cal U}$ of Ref.~\cite{Rossner} is fitted 
to LQCD data in the pure gauge theory at finite $T$~\cite{Boyd,Kaczmarek}: 
\begin{align}
&{\cal U} = T^4 \Bigl[-\frac{a(T)}{2} {\Phi}^*\Phi
	+ b(T)\ln(1 - 6{\Phi\Phi^*}  + 4(\Phi^3+{\Phi^*}^3)
            - 3(\Phi\Phi^*)^2 )\Bigr], \label{eq:E13}\\
&~~~~~~~a(T) = a_0 + a_1\Bigl(\frac{T_0}{T}\Bigr)
                 + a_2\Bigl(\frac{T_0}{T}\Bigr)^2,
 ~~~b(T)=b_3\Bigl(\frac{T_0}{T}\Bigr)^3  \label{eq:E14} .
\end{align}
The parameters included in ${\cal U}$ are summarized in Table I.  
The Polyakov potential yields a first-order deconfinement phase transition at 
$T=T_0$ in the pure gauge theory. 
The original value of $T_0$ is $270$ MeV evaluated by the pure gauge lattice QCD calculation. 
However, the PNJL model with this value of $T_0$ yields a larger value 
of the transition temperature at zero chemical potential than 
the full LQCD simulation~\cite{Karsch3,Karsch4,Kaczmarek2,Fodor} predicts. 
Therefore, we rescale $T_0$ to 212~MeV~\cite{Sakai2} so that 
the PNJL model can reproduce the critical temperature 173~MeV
of the full LQCD simulation. 

\begin{table}[h]
\begin{center}
\begin{tabular}{llllll}
\hline
~~~~~$a_0$~~~~~&~~~~~$a_1$~~~~~&~~~~~$a_2$~~~~~&~~~~~$b_3$~~~~~
\\
\hline
~~~~3.51 &~~~~-2.47 &~~~~15.2 &~~~~-1.75\\
\hline
\end{tabular}
\caption{
Summary of the parameter set in the Polyakov-potential sector
used in Ref.~\cite{Rossner}. 
All parameters are dimensionless. 
}
\end{center}
\end{table}

The variables $X=\Phi$, ${\Phi}^*$, $\sigma$, $\pi_i$, $\eta$ and $a_i$
satisfy the stationary conditions, 
\begin{eqnarray}
\partial \Omega/\partial X=0. 
\label{eq:SC}
\label{condition}
\end{eqnarray}
The solutions of the stationary conditions do not give 
the global minimum $\Omega$ 
necessarily. 
There is a possibility that they yield a local minimum or even a maximum. 
We then have checked that the solutions yield the global minimum when the solutions $X(T,\theta ,\mu )$ are inserted into (\ref{eq:E12}). 
For $\theta=0$ and $\pi$, the thermodynamic potential $\Omega$ 
is invariant under $P$ transformation, 
\bea
\eta \to -\eta,~~~~\pi_a \to -\pi_a. 
\eea

Since the four-quark coupling constant $G$ contains effects of gluons, 
$G$ may depend on $\Phi$. 
In fact, recent calculations~\cite{Braun,Kondo,Herbst} of 
the exact renormalization group equation (ERGE)~\cite{Wetterich} suggest 
that the higher-order mixing interaction is induced by ERGE. 
It is highly expected that the functional form and the strength of 
the entanglement vertex $G(\Phi)$ are determined in future 
by these theoretical approaches. 
In Ref.~\cite{Sakai5}, we assumed the following form 
\begin{eqnarray}
G(\Phi)=G[1-\alpha_1\Phi\Phi^*-\alpha_2(\Phi^3+\Phi^{*3})] 
\label{entanglement-vertex}
\end{eqnarray}
by respecting the chiral symmetry, 
$P$ symmetry, $C$ symmetry~\cite{Kouno,Kouno_CP} and the extended 
$\mathbb{Z}_3$ symmetry. 
This model is called the entanglement PNJL (EPNJL) model. 
The EPNJL model with the parameter set~\cite{Sakai5}, 
$\alpha_1=\alpha_2=0.2$ and $T_0=190$~MeV, can reproduce 
LQCD data at imaginary chemical potential~\cite{FP,D'Elia,Chen34,Chen,D'Elia-iso,Cea,D'Elia-3,FP2010,Nagata,Takaishi} and 
real isospin chemical potential~\cite{Kogut} 
as well as the results at zero chemical potential~\cite{Karsch3,Karsch4,Kaczmarek2}. 
Recently, it was shown in Ref.~\cite{Gatto:2010pt} that, also under the strong magnetic field, the EPNJL model yields results consistent 
with the LQCD data~\cite{D'Elia5}. 
The EPNJL model with this parameter set is also applied to 
the present case with real $\mu$ and finite $\theta$.

\section{Numerical Results}
\label{Nmumerical}

The present PNJL model has eight condensates of quark-antiquark pair. 
However, $\vec{a}$ and $\vec{\pi}$ vanish~\cite{Boer,Boomsma,Kouno_CP}, 
since $m_{\rm u}=m_{\rm d}$ and the isospin chemical potential 
is not considered here. 
We can then concentrate ourselves on $\sigma$, $\eta$ and $\Phi$. 
We can also restrict $\theta$ in a period $0 \le \theta \le 2\pi$ 
without loss of generality.

\begin{figure}[htbp]
\begin{center}
 \includegraphics[width=0.45\textwidth]{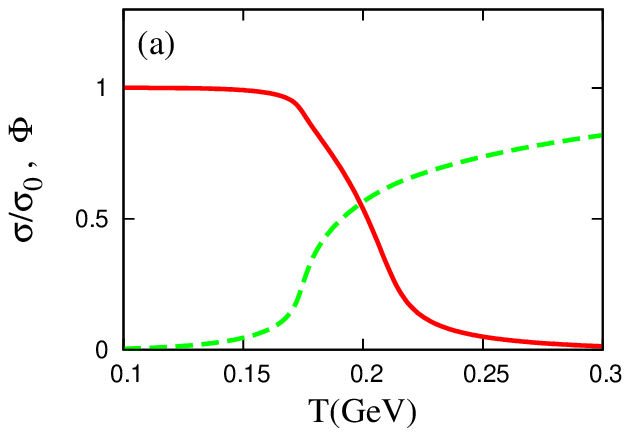}
 \includegraphics[width=0.45\textwidth]{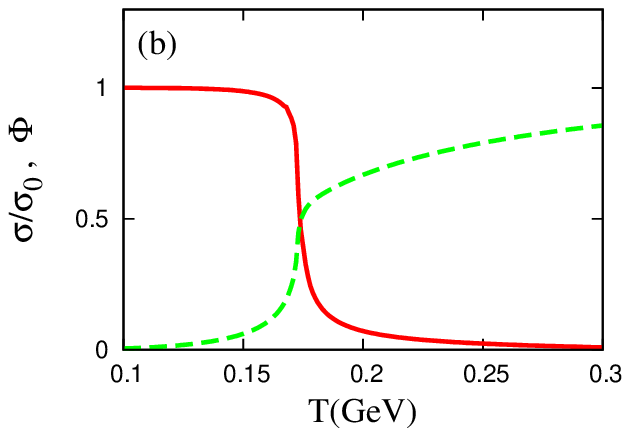}
 \end{center}
\caption{$T$ dependence of 
the chiral condensate $\sigma$ (solid line) 
and the Polyakov loop $\Phi$ (dashed line) at $\mu =0$ and $\theta =0$. 
Panel (a) represents results with PNJL model, while panel (b) does results of 
the EPNJL model. The chiral condensate is normalized 
by the value $\sigma_0$ at $T=\theta=0$.
}
\label{mu000theta000}
\end{figure}

Figure~\ref{mu000theta000} shows $T$ dependence of 
$\sigma$ and $\Phi$ at $\mu =0$ and $\theta =0$; 
note that $\eta=0$ in this case. 
In both the PNJL and EPNJL models, $\sigma$ ($\Phi$) rapidly but 
continuously decreases (increases) as $T$ increases. 
Therefore, the chiral restoration and the deconfinement transition are 
crossover. 
The crossover transitions occur more rapidly in the EPNJL model 
than in the PNJL model.

\begin{figure}[htbp]
\begin{center}
 \includegraphics[width=0.45\textwidth]{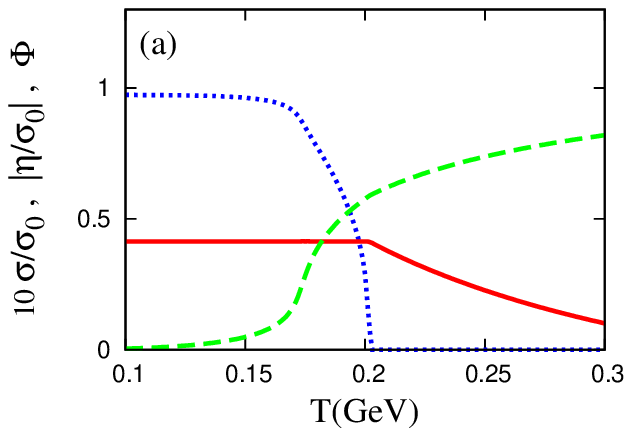}
 \includegraphics[width=0.45\textwidth]{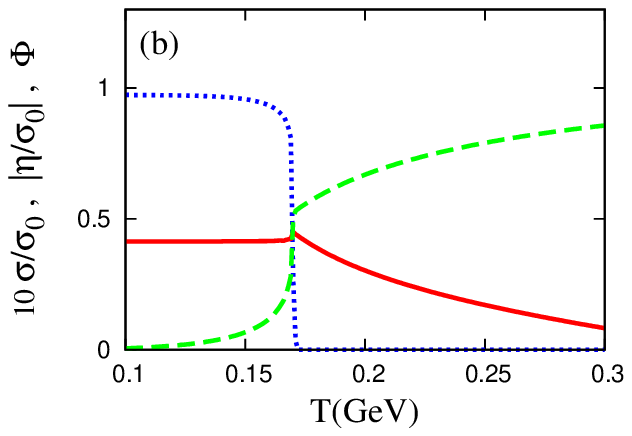}
\end{center}
\caption{$T$ dependence of the chiral condensate $\sigma$ (solid line), 
the eta condensate $\eta$ (dotted line) and 
the Polyakov loop $\Phi$ (dashed line) at $\mu =0$ and $\theta =\pi$. 
Panel(a) stands for results of the PNJL model, while panel (b) 
does results of the EPNJL model. 
}
\label{mu000theta100}
\end{figure}

Figure~\ref{mu000theta100} presents $T$ dependence of 
$\sigma$, $\eta$ and $\Phi$ at $\mu =0$ and $\theta =\pi$; 
note that this figure plots condensates $\sigma$ and $\eta$ before 
the transformation \eqref{UAtrans}. 
For both the PNJL and EPNJL models,  
the eta condensation ($\eta \neq 0$) occurs at low $T$
and hence $P$ symmetry is spontaneously broken there. 
At high $T$, in contrast, $\eta =0$ and hence $P$ symmetry is restored. 
The critical temperature $T_{P}$ of $P$ restoration is 
202 (170)~MeV in the PNJL (EPNJL) model.

The order of $P$ restoration was reported to be 
of second order in the NJL model~\cite{Boer,Boomsma} but of first order 
in the linear sigma model~\cite{Mizher}. 
In the EPNJL (PNJL) model, $\eta$ is discontinuous in its zeroth (first) 
order. 
Thus, the PNJL model supports the second-order transition, but 
the EPNJL model does the first-order. 
The zeroth-order discontinuity (gap) of $\eta$ in the EPNJL model 
is propagated to other quantities $\sigma$ and $\Phi$ as the zeroth-order 
discontinuity (gap), 
according to the discontinuity theorem 
by Barducci, Casalbuoni, Pettini and Gatto~\cite{BCPG}.
This really takes place in Fig.~\ref{mu000theta100}, 
although the discontinuity (gap) is appreciable for $\Phi$ but very tiny 
for $\sigma$. 
The first-order discontinuity (cusp) of $\eta$ in the PNJL model 
is also propagated 
to $\sigma$ and $\Phi$ as the first-order discontinuity (cusp)~\cite{Kashiwa5}. 
As mentioned above, the EPNJL model is more consistent with the LQCD data than 
the PNJL model. This means that the order of $P$ restoration may be weak first 
order.

\begin{figure}[htbp]
\begin{center}
 \includegraphics[width=0.45\textwidth]{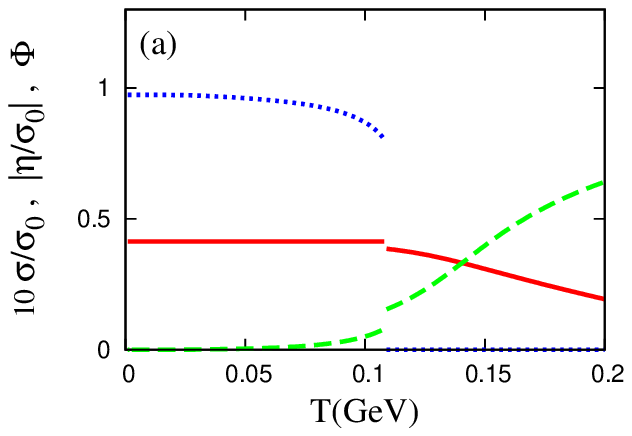}
 \includegraphics[width=0.45\textwidth]{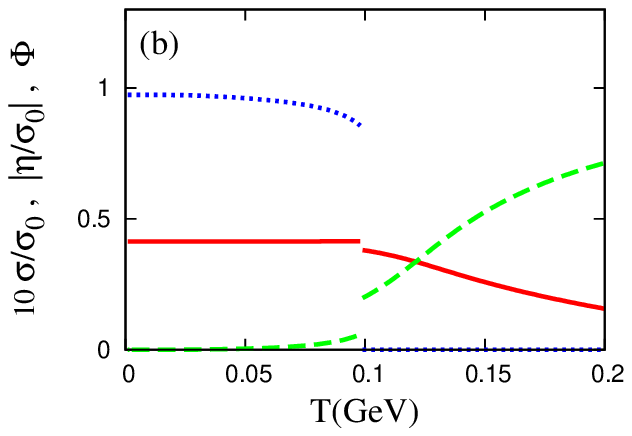}
\end{center}
\caption{$T$ dependence of the chiral condensate $\sigma$ (solid line), 
the eta condensate $\eta$ (dotted line) and 
the Polyakov loop $\Phi$ (dashed line) 
at $\mu =300$MeV and $\theta =\pi$. 
Panel(a) stands for results of the PNJL model, while panel (b) 
does results of the EPNJL model. 
}
\label{mu270theta100}
\end{figure}

$T$ dependence of $\sigma$, $\eta$  and $\Phi$ is shown also 
for $\mu =300$MeV and $\theta =\pi$ in Fig.~\ref{mu270theta100}.  
In the case of large $\mu$, $P$ restoration is of first order in both 
the PNJL and EPNJL models. 
As $\mu$ increase with $\theta$ fixed at $\pi$, thus, 
$P$ restoration changes from the second order to the first order 
in the PNJL model, while it is always of first-order 
in the EPNJL model. 
This means that there is a tricritical point (TCP) in the PNJL model; 
note that the TCP is a point 
where the first- and second-order transitions meet 
each other.

\begin{figure}[htbp]
\begin{center}
 \includegraphics[width=0.45\textwidth]{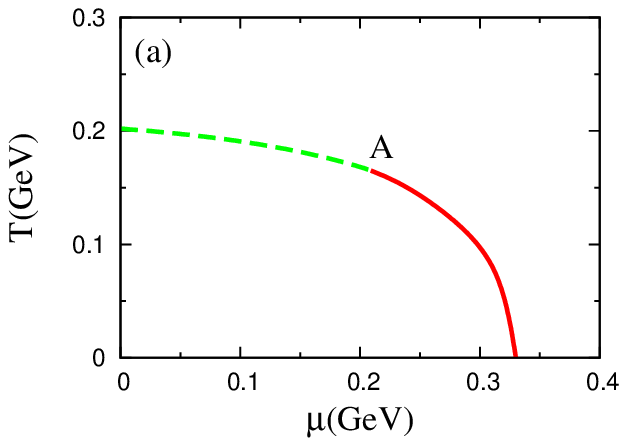}
 \includegraphics[width=0.45\textwidth]{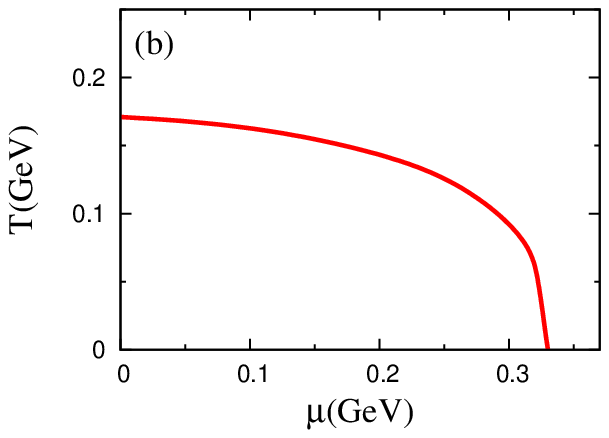}
 \end{center}
\caption{Phase diagram of $P$ restoration at $\theta=\pi$. 
Panel(a) stands for results of the PNJL model, while panel (b) 
does results of the EPNJL model. 
The solid and dashed lines represent the $P$ restoration of 
first-order and second-order, respectively.
Point A in panel (a) is a TCP
}
\label{PhaseDiagram_theta100}
\end{figure}

Figure~\ref{PhaseDiagram_theta100} shows the phase diagram of $P$ restoration 
in the $\mu$-$T$ plane for the case of $\theta =\pi$. 
In the PNJL model of panel (a), 
point A at $(\mu,T)=(209~{\rm [MeV]},165~{\rm [MeV]})$ 
is a TCP of $P$ restoration, while there is no TCP in the EPNJL model of 
panel (b).

\begin{figure}[htbp]
\begin{center}
 \includegraphics[width=0.48\textwidth]{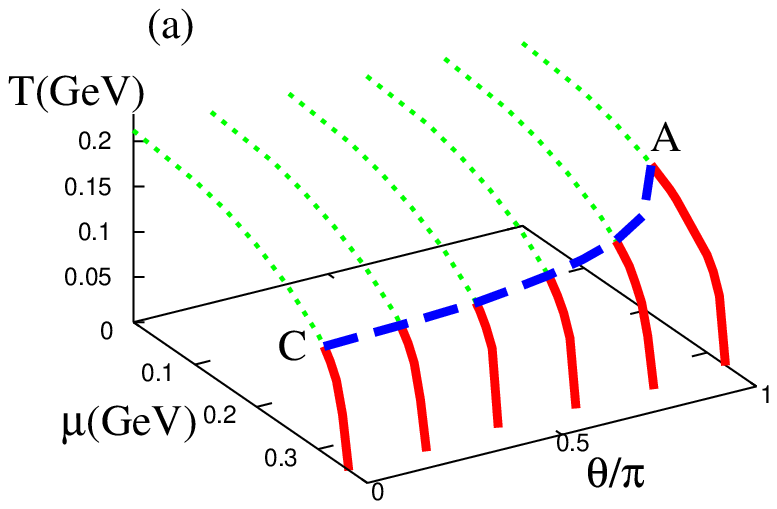}
 \includegraphics[width=0.48\textwidth]{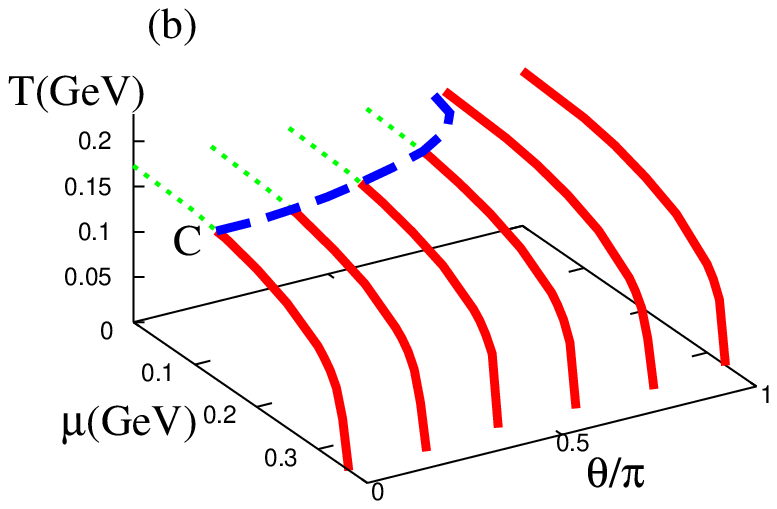}
 \end{center}
\caption{Phase diagram of the chiral restoration 
in the $\mu-\theta-T$ space. 
The solid, dashed and dotted lines represent 
the chiral transitions of first-order, second-order and crossover, 
respectively. 
Point A is a CEP of chiral restoration and a TCP of $P$ restoration, 
while point C is a CEP of chiral restoration. 
Panel(a) stands for results of the PNJL model, while panel (b) 
does results of the EPNJL model. 
}
\label{3d}
\end{figure}

Figure~\ref{3d} represents the phase diagram of the chiral 
transition in the $\mu$-$\theta$-$T$ space. 
These diagrams are mirror symmetric with respect to the $\mu$-$T$ plane 
at $\theta =\pi$.  
At finite $\theta$, $\sqrt{\sigma^2+\eta^2}$ is the order
parameter of the chiral transition rather than $\sigma$ 
itself~\cite{Kouno_CP}. 
In the PNJL model of panel (a), point A in the $\mu$-$T$ plane 
at $\theta=\pi$ is a TCP of $P$ restoration and 
a critical endpoint (CEP) of chiral restoration at which 
the first-order (solid) line is connected to the crossover (dotted) line. 
Point C in the $\mu$-$T$ plane at $\theta=0$ is another CEP of 
the chiral transition~\cite{AY,Barducci_CEP}. 
The second-order (dashed) line from C to A 
is a trajectory of CEP with respect to increasing $\theta$ from 0 to $\pi$. 
Thus,  the CEP (point C) at $\theta=0$ is a remnant of the TCP (point A) of 
$P$ restoration at $\theta=\pi$. 
In the EPNJL model of panel (b), 
no TCP and then no CEP appears in the $\mu$-$T$ plane at $\theta=\pi$. 
The second-order (dashed) line starting from point C never reaches 
the $\mu$-$T$ plane at $\theta=\pi$. 
For both the PNJL and EPNJL models, the location of CEP in the $\mu$-$T$ 
plane moves to higher $T$ and lower $\mu$ as $\theta$ increases from 0 
to $\pi$. Particularly in the EPNJL model, 
the chiral transition is always first order 
in the $\mu$-$T$ plane at $\theta=\pi$.

\begin{figure}[htbp]
\begin{center}
 \includegraphics[width=0.45\textwidth]{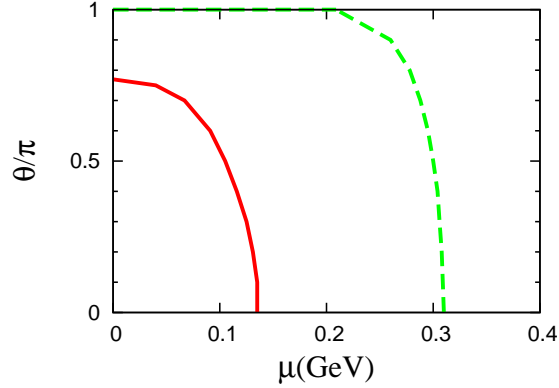}
\end{center}
\caption{Projection of the second-order chiral-transition line 
in the $\mu$-$\theta$-$T$ space on the $\mu$-$\theta$ plane. 
The solid (dashed) line corresponds to the line in the EPNJL (PNJL) model. 
}
\label{C_Plot}
\end{figure}

Figure~\ref{C_Plot} shows the projection of 
the second-order chiral-transition line in the $\mu$-$\theta$-$T$ space 
on the $\mu$-$\theta$ plane. 
The solid (dashed) line stands for the projected line 
in the EPNJL (PNJL) model. 
The first-order transition region exists on the right-hand side of 
the line, while the left-hand side corresponds to the chiral crossover 
region. The first-order transition region is much wider in the EPNJL model 
than in the PNJL model. In the EPNJL model, eventually, 
the chiral transition becomes first order even at $\mu=0$ when $\theta$ is 
large. 

QCD has the anomalous Ward identities 
among the chiral and eta condensates, $\sigma$ and $\eta$, 
the gluon condensate 
$\langle\frac{g^2}{64\pi^2}
\epsilon^{\alpha\beta\sigma\rho}F^a_{\mu\nu}F^a_{\sigma\rho}\rangle$
and the topological susceptibility $\chi_t$ \cite{MZ}:
\begin{eqnarray}
\frac{\partial\Omega}{\partial\theta}&=&\langle\frac{g^2}{64\pi^2}
\epsilon^{\alpha\beta\sigma\rho}F^a_{\mu\nu}F^a_{\sigma\rho}\rangle
=\frac{1}{N_f}m_0\eta,
\label{sus1}\\
\frac{\partial^2\Omega}{\partial\theta^2}&=&-\chi_t
=-\frac{1}{N_f^2}m_0\sigma+{\cal O}(m_0^2). 
\label{sus2}
\end{eqnarray}
The identities are useful for checking the self-consistency of 
the proposed model \cite{MZ}. 
The gluon condensate does not appear explicitly in the PNJL model, 
but the model satisfies 
\begin{eqnarray}
{\partial \Omega\over{\partial \theta}}=\frac{1}{N_f}m_0\eta.
\end{eqnarray}
The PNJL model also satisfies the second Ward identity \eqref{sus2} 
as shown below. 
In the PNJL model, the left-hand side of (\ref{sus2}) is rewritten into 
\begin{eqnarray}
\frac{\partial^2\Omega}{\partial\theta^2}
=\left(\frac{\partial^2\Omega}{\partial\theta^2}\right)_{{\rm fixed}~\phi_i}
-\sum_{i,j}\frac{\partial^2\Omega}{\partial\theta\partial\phi_i}
\left(\frac{\partial^2\Omega}{\partial\phi_i\partial\phi_j}\right)^{-1}
\frac{\partial^2\Omega}{\partial\theta\partial\phi_j}
\label{sus3}
\end{eqnarray}
with the inverse curvature matrix
\begin{eqnarray}
C^{-1}_{ij}=
\left(\frac{\partial^2\Omega}{\partial\phi_i\partial\phi_j}\right)^{-1}
\end{eqnarray}
for the parameters $\phi=(\sigma',~\eta',~\Phi,~\Phi^*)$. 
Note that $C^{-1}_{ij}$ is the susceptibility $\chi_{ij}$ of order parameters 
$\phi_i$ and $\phi_j$ and the term including $\chi_{ij}$ is of order 
$O(m_0^2)$. 
Equation (\ref{sus3}) turns out to be 
the second Ward identity (\ref{sus2}) after simple algebraic calculations. 
For $\theta=0$, Eq. \eqref{sus3} is further rewritten into  
\begin{eqnarray}
\frac{\partial^2\Omega}{\partial\theta^2}
=-\frac{1}{4}m_0\sigma
+{1\over{4}}\frac{\Omega^0_{NN}}{(2G_{-}\Omega_{NN}^0+1)}m_0^2,
\label{sus4}
\end{eqnarray}
in the virtue of the $\theta$-reflection symmetry, 
where $\Omega^0_{NN}=\frac{\partial^2\Omega^0}{\partial N^2}$ for 
the thermodynamic potential $\Omega^0$ with no 
the meson potential $U$ and no the Polyakov potential ${\cal U}$. 
\begin{figure}[htbp]
\begin{center}
 \includegraphics[width=0.45\textwidth]{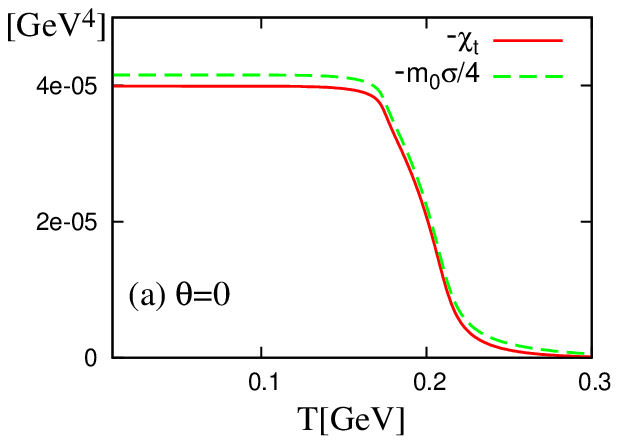}
 \includegraphics[width=0.45\textwidth]{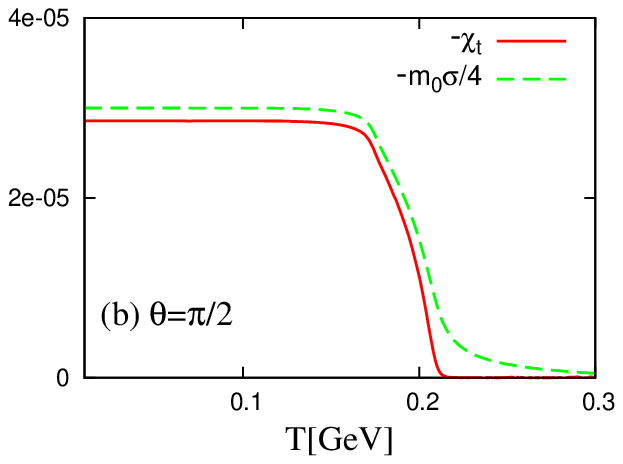}
 \includegraphics[width=0.45\textwidth]{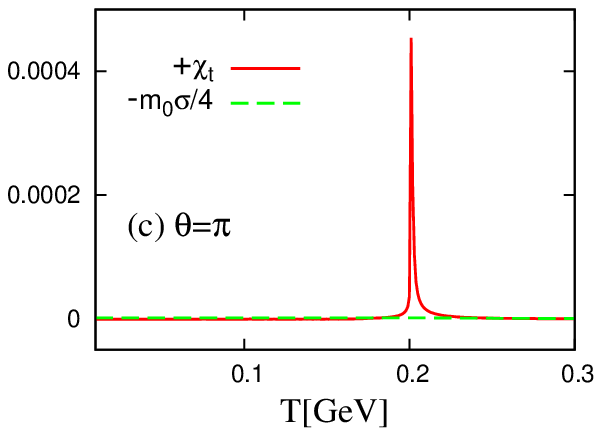}
\end{center}
\caption{$T$ dependence of the topological susceptibility $\chi_t$ 
(solid line) and the chiral condensate $\sigma$ (dashed line) 
 at (a) $\theta=0$, (b) $\pi/2$, and (c) $\pi$. 
These are results of the PNJL model at $\mu=0$. 
}
\label{TopSus}
\end{figure}

Figure \ref{TopSus} shows the topological susceptibility $\chi_t$ and 
the quantity $m_0 \sigma/4$ as a function of $T$ 
at $\theta=0,~\pi/2$ and $\pi$, 
where the case of $\mu=0$ is considered. 
As shown in panels (a) and (b), 
the topological susceptibility almost agrees with $m_0 \sigma/4$
for $\theta \le \pi/2$. 
The small deviation between the two quantities shows that 
the corrections of order $O(m_0^2)$ to the approximate identity 
$\chi_t = \frac{1}{N_f^2}m_0\sigma$ are small. 
In panel (c) for $\theta = \pi$, 
the both quantities are still close to each other 
except for the critical temperature $T \approx 200$~MeV 
of the second-order $P$ transition. 
Near the critical temperature, 
the susceptibility $\chi_{ij}$ of the order parameters becomes large 
and hence the corrections of order $O(m_0^2)$ 
coming from the second term of 
the right-hand side of (\ref{sus3}) become significant.

\section{Summary}
\label{Summary}

In summary, we have investigated theta-vacuum effects on 
QCD phase diagram, using the PNJL and EPNJL models. 
For the both models, the chiral transition becomes strong 
as $\theta$ increases. 
Particularly in the EPNJL model that is more reliable 
than the PNJL model, it becomes first order even at $\mu=0$ 
when $\theta$ is large. 
This is an important result. 
If the chiral transition becomes first order at zero 
$\mu$, it will change the scenario of cosmological evolution. 
For example, the first-order transition allows us 
to think the inhomogeneous Big-Bang nucleosynthesis model or 
a new scenario of baryogenesis. 
Our analyses are based on the two-flavor PNJL (EPNJL) model 
and effects of the strange quark are then neglected. 
It is very interesting to study the effects 
by using the three-flavor PNJL model~\cite{Matsumoto}.

\begin{acknowledgments}
The authors thank T. Inagaki, A. Nakamura, T. Saito, K. Nagata and K. Kashiwa for useful discussions and suggestions. 
H. K. also thanks M. Imachi, H. Yoneyama, H. Aoki and M. Tachibana for useful discussions and suggestions. 
Y.S. and T.S. are supported by JSPS. 
\end{acknowledgments}


\end{document}